\documentclass[12pt]{article}
\usepackage[all]{xy}
\usepackage{amssymb} 
\usepackage{amsmath} 
\usepackage{color} 
\font\frak=eufm10 scaled\magstep1
\def\goth #1{\hbox{{\frak #1}}}
\def\<#1>{\langle#1\rangle}

\def\R{\mathbb{R}}
\def\pd#1#2{\frac{\partial#1}{\partial#2}}

\def\dfrac{\displaystyle\frac}

\def\wr{{\rm w}}
\def\X{{\goth X}}
\def\V#1{\overrightarrow{\kern-2pt#1\kern0pt}}  
\def\x{\times}

\def\a{{\alpha}}
\def\b{{\beta}}
\def\l{{\lambda}}

\def\G{{\Gamma}}

\def\f{\varphi}
\def\w{\omega}
\def\W{\Omega}
\def\cinf#1{C^\infty#1}

\def\be{\begin{equation}}
\def\ee{\end{equation}}
\def\ba{\begin{eqnarray}}
\def\ea{\end{eqnarray}}
\def\ben{\begin{equation*}}
\def\een{\end{equation*}}
\def\ban{\begin{eqnarray*}}
\def\ean{\end{eqnarray*}}

\newcommand{\T}{\mathbf{T}}

\newcommand{\dg}{\mathrm{div}\,}

\catcode`@=11 \@addtoreset{equation}{section}

\newtheorem{definition}{Definition}
\newtheorem{theorem}{Theorem}

\begin{document}

\title{Some Applications of Affine in Velocities Lagrangians in Two-dimensional  Systems}

\author{Jos{\'e}\ F.\ Cari{\~n}ena\\
\small Departamento de F\'{\i}sica Te\'orica and IUMA,
Universidad de Zaragoza.\\ 
\small  50009, Zaragoza, Spain\\
\small  e-mail: jfc@unizar.es \\
\and
Jos{\'e}\ Fern\'andez-N\'u\~nez\\
\small Departamento~de F\'\i sica, Universidad de Oviedo.\\ 
\small 33007 Oviedo, Spain\\
\small e-mail: nonius@uniovi.es}
\medskip

\maketitle


\begin{abstract}
The 2-dimensional inverse problem for first-order systems is analysed and  a method to construct an affine Lagrangian for such  systems is developed.
The determination of such  Lagrangians is  based on the theory of the Jacobi multiplier
 for  the system of  differential equations. We illustrate our analysis with several examples of families of forces that are relevant in mechanics, on one side,  and of some 
 relevant biological systems, on the other.
 \end{abstract}
\bigskip

\medskip
 
\medskip

 \medskip

\pagebreak

\section{Introduction}
\indent 

The time evolution of a system is  usually described in classical mechanics by means of systems of second-order ordinary
 differential equations. Sometimes these equations can be derived from a variational principle, and the solution of the inverse problem in classical mechanics \cite{MC81,do41,ho81,WS85,STP}, i.e. finding the corresponding variational  (or Lagrangian)  description, requires then the use of a (regular) Lagrangian function $L$ depending up to  the first-order  velocities,    that is, $L=L(t,q,\dot q)$, the   $\dot q$-dependence normally being at most quadratic (included in a kinetic energy term),    although there are Lagrangians of other forms, known as non-standard Lagrangians \cite{ca05,cies10,EN12,mu08,ST13}. In any case,   the corresponding  Euler-Lagrange equations are of the desired second-order type except when $L$ is linear or,   more generally,   affine in the velocities \cite{ca03,ca88}, for which actually the equations  are of the first-order type.

Although it may seem that  the class of  Lagrangians that are affine in the velocities scarcely has interest in mechanics, 
this is not the case, see e.g. \cite{J98}, and were considered long time ago in the framework of classical field theory \cite{nb55}. Moreover, every system of second-order differential equations can be  converted into   a related one of  first-order equations  by doubling the number of independent variables.  Consequently, one can think   about the possibility   of having such a description in terms of   affine in the velocities Lagrangians  for a given mechanical system. The problem is now that we have no    basic principles from  which to derive the affine in velocities Lagrangian (or the corresponding action integral). The only way  is to consider it as    an `inverse variational problem' of the system of first-order differential equations, and ask whether given a mechanical system    with  a system of  second-order   differential equations, one can find  an affine in velocities Lagrangian whose Euler-Lagrange   equations are equivalent  in the above sense   to the given system of second-order differential equations. Actually this motivated a geometric analysis of this kind of Lagrangians in the framework of autonomous systems \cite{ca88}, with the aim of studying the inverse problem of Lagrangian mechanics
\cite{MC81,do41,ho81,WS85} and the theory of non-point symmetries. Almost simultaneously, Faddeev and Jackiw developed a
method for the quantization of such singular systems which soon became very popular and received much   from many theoretical physicists. This procedure of dealing with such systems is usually referred to as Faddeev-Jackiw (FJ) quantization  method \cite{BNW92b,FJ88,KMK91}.

Obviously our interest can be extended to include arbitrary systems of first-order differential equations, not only those coming from  systems of second-order differential equations, because systems of first-order differential equations  play a relevant r\^ole in many cases, not only in physics, where many equations are first-order, as in the Dirac equation, but
also in other fields as  biology dynamics \cite{fn98,Tru74}, economy  and chemistry. In this article we only consider  the simplest case, the 2-dimensional one, that is, the inverse problem of the first-order system on $\R^2$ for such Lagrangians. 

The search for Lagrangians for a single second-order differential equation has received much interest during the last years, 
see \cite{CIMM,ca05,cies10,mu08,nu10,SK17,TPCSL}. The main reason for that interest is that the knowledge of  first-integrals  is very useful to integrate the system and the analysis of the infinitesimal point  symmetries of the Lagrangian can be used, via the well known  Noether theorem, to find  first-integrals. 
The study of the integrability, as well as linearisability,  of a given system is therefore  usually based on the existence of such constants of motion (see e.g.  the series of papers \cite{CSL05,CSL06,CSL09a,CSL09b,CSL09c}).  Another reason is the search for a Hamiltonian formulation in order to proceed to its possible quantization.  We therefore believe that a first-order approach to  this problem may also  be  of interest. The general treatment of the variational inverse problem for systems of first-order equations began   with the work of Havas 	\cite{ha73}. Santilli also dealt in  \cite{sa83}  with this subject and  developed   what he called  Birkhoffian mechanics \cite{B27,Z20},   that he considered as a generalisation of  Hamiltonian formalism; see also \cite{ho81} and references therein. More recent works on a   geometric approach to the problem are \cite{ca93} and \cite{ca03}. 
    
The plan of the paper is the following. In  Section 2 the main features of a 2-dimensional affine in velocities Lagrangian are analysed. In order to make the paper more self-contained and to fix notation   we 
summarise  some concepts  used in the modern geometrical formulation of mechanics (operations with vector fields and differential forms), which can be found in  classical textbooks (see e.g. \cite{ab81,CIMM,cra86}). In Section 3, although we start with a time-dependent dynamical system in $\R\x T\R^2$ 
of the Lagrangian type, it is shown that we finally have to deal with a time-dependent dynamical system  $\G\in\X(\R\x\R^2)$ which actually  is  Hamiltonian with respect to an appropriate Hamiltonian structure.
Section 4 is devoted to analyse the two-dimensional inverse problem for systems of first-order differential equations. As the main result is to be expressed in terms of a Jacobi multiplier  of the given system,  we first review in Subsection 4.1  the theory of Jacobi multipliers in  geometrical terms \cite{AFGG,AP21,CFN21,CLR15,CR21,cs21,CLS,cras08,GGLM,GGK,GG,GG22,GG11,GC13,GC15,Cl09,CN05,NL08c,nu10}. The main result asserts that in order to have a  Lagrangian description for a given system of two first-order differential equations it is necessary and sufficient to establish   the existence of  a Jacobi multiplier for the system. 
Moreover, these Jacobi multiplier can be used to find constants of motion via Hojman symmetries \cite{CR21}. The equations determining  the multipliers have always a local solution, so this inverse   problem has a positive answer; in fact, there are infinitely many (not necessarily   gauge-equivalent)
Lagrangians  for the given system, opening in this way the possibility of finding constants of motion (see e.g. \cite{CFN21,CN05,NL08c}) and alternative Lagrangians \cite{CI83}.  For more information on the r\^ole of Jacobi multipliers in integrability and with 
other approaches to integrability see e.g \cite{MCSL14,MCSL15,MCSL16,MCSL19}.  In Section 5  
we exhibit explicit Lagrangians for some important examples of differential equations, as mechanical systems, and interesting results on biological examples \cite{GGG}, as  a  generalisation of the Lotka-Volterra model \cite{CD22,fn98,VV31} and a host-parasite model \cite{Tru74},  are derived 
 from this new perspective and their  Hamiltonian functions and a set of canonical variables are also given. Finally in Section 6 we summarise the previous results and the
  generalisation to systems involving more variables is proposed.

\section{Affine Lagrangians on $\R^2$}
\indent

The  higher order in the velocities terms of the functions  usually  appearing as Lagrangians of  classical mechanical systems  are quadratic, and the 
corresponding systems of Euler-Lagrange equations are systems of second-order differential equations as it happens for more  general regular Lagrangians.
In Physics it is also frequent to use singular Lagrangians, and constraints are present. An instance of singular Lagrangians are those depending  linearly (or, more generally, affinely) on the velocities which give rise to systems of first-order 
differential equations (see for instance  \cite{ca03,ca88} and references therein). 

In order to analyse  the structure of this kind of Lagrangians, let us consider a general time-dependent Lagrangian system  on a configuration space $\R^2$ described by a Lagrangian function $L\in\cinf(\R\x T\R^2)$ which 
is affine in the velocities. In the local coordinates $(t,x,y)$ in $\R\x\R^2$ and the corresponding fibred ones  $(t,x,y,v_x,v_y)$ in  $\R\x T\R^2$ such a Lagrangian is of the form  
\be
\label{ala}
L(t,x,y,v_x,v_y)=m_x(t,x,y)\,v_x+m_y(t,x,y)\,v_y+H(t,x,y),
\ee
where $m_x,m_y,H\in\cinf(\R\x\R^2)$. Such a  Lagrangian (\ref{ala}) can be intrinsically defined  by the 1-form  on $\R\x\R^2$
\begin{equation}
\lambda=m_x(t,x,y)\,dx+m_y(t,x,y)\,dy+H(t,x,y)\,dt\label{1flambda}
\end{equation} 
by contraction with the total-time  derivative operator  
$$
\T=\pd{}t+v_x\pd{} x+v_y\pd{}y,
$$   
that is,  $L=i_\T\lambda$, where $dt\in\bigwedge\!^1(\R\x\R^2)$ 
is the pull-back of the volume  form $dt\in\bigwedge^1(\R)$ such  that $i(d/dt)dt=1$. Note that $\T$ is a vector field along the natural projection $\pi:\R\x T\R^2\to \R\x\R^2$,   given by $\pi(t,x,y,v_x,v_y)=(t,x,y)$, while $\lambda$ is the above mentioned  1-form (\ref{1flambda}) on $\R\x\R^2$; the contraction $i_\T\lambda$     makes sense and it is a real function on the manifold $\R\x T\R^2$. See \cite{ca93}, \cite{ca03},  and references therein for more details.

The system of Euler-Lagrange equations corresponding to the function  
$L\in C^\infty( \R\x T\R^2)$ given by  (\ref{ala}) is a system of two first-order differential equations that can be written in matrix form as
\be
\label{ecl}
M\dot X=W,
\ee
with the matrices  
$$M=\left(\begin{array}{cc} 0&\mu\\-\mu&0\end{array}\right)\,,\qquad 
X=\left(\begin{array}{c} x\\ y\end{array}\right)\qquad {\rm and}\qquad  
W=\left(\begin{array}{c} \wr_x\\ \wr_y\end{array}\right)
$$ 
being given by
\be
\label{mu}
\mu=\pd{m_y}{x}-\pd{m_x}{y},\quad \wr_x=\pd{m_x}{t}-\pd{H}{x},\quad
  \wr_y=\pd{m_y}{t}-\pd{H}{y}.
\ee

Observe that the functions $\mu$, $\wr_x$, and $\wr_y$ are not  completely independent because  we have, as a 
direct consequence of definition (\ref{mu}), 
\be
\label{mu/w}
\pd{\mu}{t}+\pd{(-\wr_y)}{x}+\pd{\wr_x}{y}=0.
\ee
Moreover, taking into account  that the 2-form $d\lambda$ is given by 
\be
d\l=\wr_x\,dt\wedge dx+\wr_y\,dt\wedge dy+\mu\,dx\wedge dy,
\label{diflambda}
\ee
 the relation (\ref{mu/w})  expresses nothing but  that $d^2\l=0$.

Obviously the only case of interest is the  {\sl regular} one, namely the case $\mu\ne0$ everywhere; then $M$ is invertible 
and consequently the system (\ref{ecl}) reads $\dot X=M^{-1}W$, that is
\be
\label{ec1}
\left\{
\begin{split}
\,\dot x&=-{\dfrac{\wr_y}{\mu}}\\
\dot y&={\dfrac{\wr_x}{\mu}}
\end{split}
\right.\,.
\ee
In summary, the system of Euler--Lagrange equations of a Lagrangian $L$ affine in the velocities  turns out to be  a system of first-order differential equations in normal form. 

Let us analyse the  above result in geometric terms. The system of differential equations (\ref{ec1}) determines the integral curves of the vector field
\be
\label{vfg}
\G=\pd{}{t}-\frac{\wr_y}{\mu} \pd{}{x}+\frac{\wr_x}{\mu}\pd{}{y}\in\X(\R\x\R^2).
\ee
This vector field and the 2-form $d\lambda$ on $\R\x \R^2$ enjoy the following important properties:

\medskip
1. The closed 2-form $d\l$ on $\R\x \R^2$ has rank 2, because if $X=C\,\partial_t+ A\,\partial_x+B\,\partial_y$, where $A,B,C\in C^\infty(\R\x\R^2)$,  is in the kernel of $d\lambda$,  $i(X)\,d\lambda =0$, 
as we assumed that $\mu\ne 0$ in each point, then,
$$
B=C\,\frac{\wr_x}\mu,\quad A=-C\,\frac {\wr_y}\mu\,,
$$
and  therefore, $X=C\,\Gamma$. Thus, the vector field   $\G$ is the only vector field such that $i(\G)dt=1$ and $i(\G)d\lambda=0$; the first one of these two conditions, $i(\G)dt=1$,  means that the time coordinate $t$ is the parameter
for the integral curves of $\G$, while the second equation is the intrinsic expression of the 
Euler-Lagrange equations (\ref{ecl}). Note also that as the dynamics $\Gamma$ is only determined, up to reparametrization,  by 
$d\lambda$ and not by $\lambda$, one can add to $\lambda$ any closed 1-form $\alpha \in \bigwedge\!^1(\R\x\R^2)$,
and therefore we can change $\l$ by $\l'=\l+df$, with $f\in\cinf(\R\x\R^2)$;  that is to say, the dynamics is invariant under the gauge transformation
\begin{equation}
m_x\mapsto m_x+\partial_xf,\ m_y\mapsto m_y+\partial_yf,\  H\mapsto  H+\partial_tf. \label{dfamb}
\end{equation}
 
\medskip

2. Let $\w_0=dx\wedge dy\in\bigwedge\!^2(\R^2)$ be the `natural' symplectic structure  on $\R^2$ and $\W=dt\wedge dx\wedge dy$ be the induced volume form  on $\R\x\R^2$ (in the Cartesian coordinates $(t,x,y)$). The 3-form $dt\wedge d\l$ is a volume form too and therefore  it is proportional to $\W$; actually, from the expression (\ref{diflambda}) we see that $dt\wedge d\l=\mu\,\W$, and  by contracting with the vector field $\G$ 
both sides of this last  relation we immediately get $\mu\,i(\G)\W=d\lambda$. That is to say, although the differential 2-form $\a_\G=i(\G)\W$  is not closed, the 2-form $\b_\G=\mu\,\a_\G$ is exact, $\beta_\G=d\l$. Note also that the Lie derivative 
$\pounds_{\mu\G}\W$ vanishes;
as we will see later on (Section 4), such a  property of $\mu$, which is equivalent to $\pounds_{\G}(\mu\W)=0$,  defines the so  called    `Jacobi  multipliers' for the vector field $\G$ with respect to the volume form $\Omega$. 
    
 \medskip
 
{\bf Remark 1.}    It is worth noting that although we have started with a dynamical system on $\R\x T\R^2$ of the Lagrangian type, we arrive 
at an equivalent reduced dynamics  on $\R\x\R^2$ given by the vector field (\ref{vfg}). The Lagrangian dynamics  on $\R\x T\R^2$ is given by
 the integral curves of the {\it first prolongation} $\G^1$ of $\G$, i.e., the first prolongation of the integral curves of $\G$. 
In this approach, the affine Lagrangian (\ref{ala}) is singular, and, consequently, it is subjected to constraints. In fact, the constraints are 
exactly the equations of motion (\ref{ecl}). We do not insist on this aspect because more  details can be found  in \cite{ca03}.

\section{Hamiltonian formulation}
\indent 

Now we can analyse whether there exists  a, in general time-dependent, Hamiltonian formulation for
the dynamical system $\G\in\X(\R\x\R^2)$ 
describing the dynamics of our Lagrangian (\ref{ala}) in the regular case $\mu(t,x,y)\ne0$. The problem is to look for a pair $(\w,\widetilde H)$, where
$\w\in\bigwedge\!^2(\R\x\R^2)$ is a rank two closed 2-form    and  the function $\widetilde H\in\cinf(\R\x\R^2)$ is  
such that $i(\G)\w_{\widetilde H}=0$, where $\w_{\widetilde H}=\w+d\widetilde H\wedge dt$ \cite{ab81,VP92}. For a given $\G$ the pair $(\w, \widetilde H)$ may be  not unique. When $\w$ is written in Darboux coordinates, i.e. $\w=dq\wedge dp$, 
we have a Hamiltonian description with canonical conjugate variables $q,p$ and Hamiltonian function $\widetilde H$.

As we already know, $i(\G)dt=1$ and $i(\G)d\l=0$, so the problem may be reduced to see whether $d\l$ can be put in the  needed form $\w+d\widetilde H\wedge dt$.  
Recall that $d\l$ given by (\ref{diflambda}) is also a rank two closed 2-form, as indicated above,  and that by making  use of the definition (\ref{1flambda}) it can be expressed as
\be
d\l=\beta_0+dH\wedge dt,
\ee 
 where $\beta_0$ is the 2-form $\b_0=dm_x\wedge dx+dm_y\wedge dy$,  
which is also  closed and has rank two, as it is straightforward to check. Therefore, the solution of the Hamiltonisation problem is achieved by choosing $\w=\b_0$ and $\widetilde H=H$, so that $\w_{\widetilde H}=\w_{H}=d\l$.

In practical cases, the simplest solution is obtained when considering that we can impose restrictive conditions to the
 components of the 1-form $\l$ without altering the dynamics:  as we know, the 1-form $\l'=\l+df$, $f\in\cinf(\R\x\R^2)$, is gauge-equivalent
  to $\l$; when choosing $f$ in such a way that $m_y+\partial_yf=0$, we have $\l'=(m_x+\partial_xf)dx+(H+\partial_tf)dt$. 
  For this reason, the reduced form $\l=m_xdx+Hdt$ can always be used and then $\w_H=d\l=dx\wedge d(-m_x)+dH\wedge dt$. 
  That means that a possible set of conjugate canonical variables is $q=x,\, p=-m_x$, with Hamiltonian function $H$.

\medskip
 
{\bf Remark 2.}
The  formulation we have obtained is not the Hamiltonian counterpart (via the Legendre transformation)  on $\R\x T^*\R^2$ 
of the singular Lagrangian  (\ref{ala}) we have started with; the dynamical system (\ref{vfg}) is a time-dependent 
regular Hamiltonian system $(d\l,H)$ on $\R\x\R^2$.

\section{The inverse problem for first-order systems}
\indent 

We have shown that an affine Lagrangian (\ref{ala}) gives rise to the system of  first-order differential equations (\ref{ec1}) and we immediately ask for the inverse problem; namely, given a system of ordinary first-order differential equations, is there a Lagrangian of type (\ref{ala}) whose associated Euler-Lagrange equations are equivalent to that system?  In this section we analyse this problem and conclude that, in general, it has always a positive answer; moreover,  there are infinitely many affine Lagrangians for a given first-order system. The main result is expressed in terms of  the so called Jacobi multipliers \cite{cras08,Cl09, CN05, NL08c,nu10}, a notion which we present in geometrical terms \cite{CFN21,cs21}.

\subsection{Theory of the Jacobi multipliers}
\indent

 Although we present the notion and the basic properties of  Jacobi multipliers in the  case of $\R^3$, the  constructions can be  easily extended to a multidimensional oriented manifold.

\begin{definition}
We say that the nonvanishing function $\mu\in C^\infty(\R\x\R^2)$ on the oriented manifold $(\R\x\R^2,\Omega)$ is a {\it Jacobi multiplier}  for the vector field $V\in\X(\R\x\R^2)$ with respect to the volume form $\Omega$ if   the 2-form $\b_V=\mu\, i(V)\W$ is closed. That is, $\b_V$ is locally exact, what means that there exists a locally defined  1-form  $\l\in\bigwedge\!^1(\R\x\R^2)$ such that $\b_V=d\lambda$.
\end{definition}

Note that as $\W$ and $\mu\,\W$ are volume forms, and then $d\W=d(\mu\,\W)=0$, 
the condition defining a Jacobi multiplier can be expressed in terms
 of Lie derivatives either as $\pounds_V(\mu\,\W)=0$ or as $\pounds_{\mu V}\W=0$, because $\pounds_V(\mu\,\W)=d(\mu\,i(V)\Omega)=\pounds_{\mu V}\W$,  that is,
\be
\label{jm}
\dg(\mu \,V)=0,
\ee
where ``div'' stands for the divergence operator on vector fields associated to the volume form $\W$, which is defined by
$\pounds_V\W=(\dg V)\W$,  with $V\in\X(\R\x\R^2)$ (see for instance  \cite{cra86}).

In the Cartesian coordinates $(t,x,y)$ of $\R\x\R^2$, if $\W$ is such that $\W=dt\wedge dx\wedge dy$ 
and 
\be
\label{gvf}
V=V_t\,\pd{}{t}+V_x\,\pd{}{x}+V_y\,\pd{}{y}
\ee
we recover the familiar expression  for the divergence operator of a vector field 
\be
\dg V=\pd{V_t}{t}+\pd{V_x}{x}+\pd{V_y}{y}.
\ee
Thus, if the  vector field $V\in\X(\R\x\R^2)$ is  such that $V_t=i(V)dt=1$, 
the equation (\ref{jm}) for its Jacobi multipliers $\mu$  is the partial differential equation
\be
\label{mun}
\dg(\mu\,V)=\pd{\mu}{t}+\pd{(\mu\, V_x)}{x}+\pd{(\mu\, V_y)}{y}=0.
\ee

Taking into account the properties of the div-operator, in particular $\dg(fV)=f\,\dg V+V(f)$, because 
$$
\pounds_{fV}\Omega=\pounds_{V}(f\Omega)=V(f)\Omega+f\pounds_{V}\Omega=\big(V(f)+ f\,\dg V\big)\Omega,
$$
and the fact that $V(\log f)=V(f)/f$ 
(with $f$ a positive  function), we see that the relation (\ref{jm})  between $\mu$ and $V$ can be written  as
\be\label{jmeq}
V(\mu)+\mu\,\dg V=0, 
\ee
or equivalently as
\be 
\label{logmu}
V(\log\mu)+\dg V=0.
\ee

From here, one can immediately derive several properties of the multipliers:

1) The Jacobi multipliers for divergence-free vector fields  are its  first-integrals. 

2) The multiplier for a vector field $V$ is not unique, every two multipliers $\mu$ and $\mu'$ being related by $\mu'=f\,\mu$, 
where $f$ is a  first-integral of the vector field $V$ (in the cases of interest $f$ is non-trivial); the corresponding exact 2-forms
 $d\l$ and $d\l'$ are related in the same way, $d\l'=f\,d\l$.

3)  Given  a   function $R$  such that $V(R)=\dg V$, then, having in mind that $V(e^{-R})=-e^{-R}V(R)$, one sees that  $\mu=e^{-R}$ is a Jacobi multiplier for $V$. 
For instance, in the particular case of $\dg V=\nu(t)$,  the function 
$\mu(t)=\exp\left(- {\displaystyle \int^t} \nu(t')dt'\right)$ is a Jacobi multiplier for $V$, because 
$V\left({\displaystyle  \int^t} \nu(t')dt'\right)=\nu(t)=\dg V$.

Finally, we point out some additional properties. As $\W=dt\wedge dx\wedge dy$ is a volume form on a 3--dimensional space, we can replace the vector field $V$ (\ref{gvf}) by the 2-form $\a_V=i(V)\W=V_t\,dx\wedge dy+V_x\,dy\wedge dt+V_y\,dt\wedge dx$. If $\mu$ is
a Jacobi multiplier,  then the contractions with $V$  of the (locally) exact 2-forms  $\a_V$ and  $\b_V=\mu\,\a_V$  vanish trivially, i.e. $i(V)\b_V=0$, what means that $\b_V$ is an absolute integral invariant \cite{ab81,CFN21}  of $V$ because also   $d\b_V=0$. 

For a vector field  such that $i(V)dt=1$, the volume form $\W$ is 
such that  $\W=dt\wedge\a_V$. To prove this identity  it is enough to observe that $\W$ and $dt\wedge\a_V$ 
are both 3-forms on a 3-dimensional manifold, so they are proportional, and, as $i(V)\alpha_V=0$,  the condition $V_t=1$ immediately yields the identity. It can also be checked  by using the corresponding  local expressions.
Moreover, the 2-form $\a_V$ also satisfies the identities $\pounds_V\W=d\a_V$ and  $\pounds_V\a_V=(\dg V)\,\a_V$, because $\pounds_V\a_V=\pounds_V(i(V)\Omega)=i(V)\pounds_V\Omega=\dg V\,\alpha_V$. 

\subsection{The  inverse problem  on $\R^2$}
\indent

Let us now consider the inverse problem on $\R^2$. Given a  non-autonomous system of ordinary first-order differential equations on $\R^2$
\be
\label{ec2}
\left\{\begin{split}
\,&\dot x=X(t,x,y)\\
&\dot y=Y(t,x,y)
\end{split}\right.\,,
\ee
we can consider an associated system 
\be
\label{ec2b}
\left\{\begin{split}
\,&\dfrac{dt}{ds}=1\\
&\dfrac {dx}{ds}=X(t,x,y)\\
&\dfrac{dy}{ds}=Y(t,x,y)
\end{split}\right.
\ee
whose solutions are the integral curves of the vector field on $\R\x\R^2$
\be
\label{nfode}
\G=\pd{}t+X(t,x,y)\pd{}x+Y(t,x,y)\pd{}y\in\X(\R\x\R^2).
\ee

The existence of a Hamiltonian description for the system was carried out in  \cite{TdC06,TdC09} and it was related to the determination of a Jacobi multiplier for the vector field (\ref{nfode}). We want to study here the inverse problem of Lagrangian mechanics  for this system. In other words, 
Is there a Lagrangian  $L(t,x,y,v_x,v_y)$ affine in the velocities whose system of Euler-Lagrange equations is equivalent to the system (\ref{ec2})? Once an affine Lagrangian has been found, we have also a Hamiltonian formulation (see  the previous section).
 The answer to this question  is given by the following theorem. 

\medskip

\begin{theorem} Given a   vector field $\G\in\X(\R\x\R^2)$  as in (\ref{nfode}),  then there exists a 1-form $\lambda\in\bigwedge\!^1(\R\x\R^2)$ defining an affine in velocities Lagrangian $L=i_\T\lambda$ giving rise to the first-order system (\ref{ec2}) if, and only if, there exists a
 Jacobi multiplier $\mu$ for $\G$ with respect to the volume form $\W=dt\wedge dx\wedge dy$ in $\R\x\R^2$. In this case, the 1-form $\l$ is  
 such that $d\l =\mu\,i(\G)\W$.
 \end{theorem}
{\sl Proof.} If $\G$ derives from the affine Lagrangian $L=i_\T\l$, where $\lambda\in \bigwedge^1(\R\x\R^2)$, then $i(\G)d\l=0$. As we have pointed out in Section 2, $dt\wedge d\l=\mu\,\W$, 
where the proportionality  factor $\mu$ is given by the first expression in (\ref{mu}). Then,  by contraction with the vector field $\G$ we immediately get $d\l=\mu\, i(\G)\W$.  Conversely,
 if $\mu$ is a Jacobi multiplier for the   vector field $\G$,  then there exists a 1-form $\l$ such that $d\l=\mu\, i(\G)\W$ and  hence  $i(\G)d\l=0$, which means that $\G$ is of the Lagrangian type, with associated 1-form $\l$, i.e. $L=i_\T\l$.

\medskip

It is clear  that every Jacobi multiplier  for $\G$ gives rise to various (gauge-equivalent)  Lagrangians. In fact, a given multiplier $\mu$ determines the cohomology class 
  of   1-forms  $\l+df$, with $f\in\cinf(\R\times\R^2)$,  and the Lagrangians  determined by $\l$ and $\l+df$ are gauge-equivalent, $L'=L+df/dt$. The 
  gauge-equivalence  imposes some restrictive conditions (\ref{1flambda}) on the coefficients $m_x$, $m_y$ and $H$.

On the other hand, given two different multipliers $\mu$ and $\mu'$ the function $f$ such that  $\mu'=f\,\mu$  is a  first-integral of $\G$, as  pointed out previously. Then $d\l'=f\,d\l$ and if $f$ is not a constant function the respective Lagrangians $L$ and $L'$ 
 are (solution-) equivalent but not gauge-equivalent Lagrangians (see e.g. \cite{ca85}).

In conclusion, there are infinitely many affine Lagrangians for the system  (\ref{ec2}) or the vector field  (\ref{nfode}), a result also reflected in the system of PDE which 
must satisfy the coefficients $m_x$, $m_y$ and $H$ (see (\ref{myH}) below).

Now, in order to find an affine Lagrangian for a vector field $\G$  given by (\ref{nfode}) we can devise a specific procedure by using the 
result of the theorem. We proceed along the following steps:

\medskip

1) Find a particular solution of the PDE corresponding to (\ref{mun}) for the Jacobi multiplier $\mu$, i.e. (\ref{mun}) with $V_x=X$ and $V_y=Y$.

2) Once such a particular solution $\mu$ has been found, we have to look for the coefficients $m_x$, $m_y$ and $H$ to build  the 1-form $\l=m_xdx+m_ydy+Hdt$ such that $d\l=\mu\, i(\G)\W$. At this 
stage we firstly construct the 2-form $\b_\G=\mu\, i(\G)\W=\mu(dx\wedge dy+Y\,dt\wedge dx-X\,dt\wedge dy)$ and then the condition $\b_\G=d\l$ gives  the system of linear PDE for these coefficients:
\be
\label{myH}
\left\{\begin{split}
\,&{\displaystyle\pd{m_y}x-\pd{m_x}{y}}=\mu\\
&{\displaystyle\pd{m_x}{t}-\pd{H}{x}}=\mu\, Y\\
&{\displaystyle\pd{m_y}{t}-\pd{H}{y}}=-\mu\, X
\end{split}\right. .
\ee
 Assuming that the needed regularity conditions are fulfilled, this system has always solution (Cauchy--Kovaleski theorem) but no general 
method  to solve it does exist, of course. 

However, we  only need one particular solution of (\ref{myH}) and the gauge-invariance property of the Lagrangian  allows us to impose additional restrictive conditions to the $m$-coefficients in order to simplify the final expression of $L$. 
In several of the following examples,  we look for a solution such that $m_x\ne0$, $m_y=0$, and $H\ne0$, as indicated at the end of Section 3. If such a solution does exist, every Lagrangian related to this Jacobi  multiplier $\mu$ is gauge-equivalent to the Lagrangian $L_1=m_xv_x+H$, i.e. $L=L_1+df/dt$ for some function $f=f(t,x,y)$.

 \section{Applications to mechanical and biological systems}
 \indent 
 
To solve an inverse problem for a given $t$-dependent vector field $\Gamma$ 
as in (\ref{nfode})  one must first look for a Jacobi multiplier $\mu$, that is, a particular solution of the equation corresponding to  (\ref{mun}) with $ V_x=X$ and $V_y=Y$ for $V=\Gamma$. There are methods for the determination of Jacobi multipliers, in particular when the vector field is of a polynomial type (see e.g. \cite{ca09a,ca11} for two recent examples).  Once a multiplier $\mu$ has been found, we have to look for a 1--form $\lambda$ such that  $d\lambda=\b_\G=\mu(dx\wedge dy+Y\,dt\wedge dx-X\,dt\wedge dy)$, which can be done by direct inspection or by solving (\ref{myH}).  Then, we can  construct the  Lagrangian and Hamiltonian descriptions according to  Sections 2 and 3.  In the following applications, the volume form $\W=dt\wedge dx\wedge dy$ is always understood.
 
 \subsection{Mechanical systems}
\indent 

  Usual mechanical systems are described by a second-order differential equation (a SODE) $\ddot x=F(t,x,\dot x)$, where $F$ is the total force. Denoting the basic coordinates by $(t,x,v)$, this differential equation has associated a system of first-order differential equations 
\begin{equation}
\left\{ \begin{split}
\, &\dot x=v\\
&\dot v=F(t,x,v)
  \end{split}\right.,
 \label{sodesys}
  \end{equation}
and  according to  (\ref{nfode}) the equivalent vector field $\G\in\X(\R\x\R^2)$ is 
\be
 \G=\pd{}t+v\pd{}{x}+F(t,x,v)\pd{}{v}\,.
 \label{mech}
 \ee
 If  $\W=dt\wedge dx\wedge dv$,  ${\rm div\,}\Gamma=\partial F/\partial v$ and the equation (\ref{logmu}) for the Jacobi last multiplier $\mu$ is 
 \be
 \label{muforce}
 \G(\log\mu)+\pd{F}{v}=0.
 \ee
 It is evident in this expression that that for $v$-independent forces the Jacobi multipliers are but the  first-integrals. Consider therefore
 as a first example  the particular case of a velocity-independent force, $F=F(t,x)$.  Thus the equation (\ref{muforce}) is $\G(\log\mu)=0$, and then any real constant is a
 Jacobi  multiplier. The simplest choice,  $\mu=1$, gives 
 \ben
 \b_\G=\alpha_\Gamma=dx\wedge dv-v\, dt\wedge dv+F(t,x)\, dt\wedge dx,
  \een
  and then, we see that
   \ben
 \beta_\G=d\left(-v\,dx-\Big(\int^xF(t,z)\,dz-\frac{v^2}{2}\Big)dt\right). 
 \een
Consequently, $\beta_\G=d\l$, with 
 \ben
 \lambda=-v\,dx+\left(\frac{v^2}{2}-\int^xF(t,z)\,dz\right)dt, 
 \een
  from where we see that $m_x$, $m_v$ and $H$  in (\ref{1flambda})  are given by 
\be
\label{mmH}
 m_x=-v,\quad m_v=0,\quad H=\frac{v^2}{2}-\int^xF(t,z)\,dz ,
\ee
and the Lagrangian $L=i_\T\lambda$ is then
 \be
 \label{lf(t,x)}
 L(t,x,v,\dot x,\dot v)=-v\,\dot x+\frac{v^2}{2}-\int^xF(t,z)\,dz.
 \ee
Here, and in all the examples of mechanical type,  the `velocities' of $x$ and $v$ are  directly denoted, as usual, by $\dot x$ and $\dot v$, respectively.

It is an easy task to check that (\ref{mmH}) is a solution of the system (\ref{myH}), where $y$ is $v$, $\mu=1$, $X=v$, and $Y=F(t,x)$. It is also easy to derive from the affine Lagrangian (\ref{lf(t,x)}) 
that we have found the system of first-order equations of motion (\ref{sodesys}) equivalent to the second-order differential equation $\ddot x=F(t,x)$. The Lagrangian energy is $E_L={\displaystyle \frac{v^2}2- \int^x}F(t,z)\,dz$,  that  in general is not a conserved quantity.

As far as the Hamiltonian formulation is concerned, from the expression of $\l$ we have $\w_H=d\l=dx\wedge dv+dH\wedge dt$, so that the set of  canonical conjugate variables is, up to a canonical transformation, 
$q=x,\, p=v$. The Hamiltonian function is written $$H=\frac{p^2}2-{\displaystyle  \int^q} F(t,z)\,dz.$$

 \medskip
 
  The inverse problem for forces that may depend on the velocities clearly plays an interesting r\^ole, as it is next illustrated. So, 
  as a second example, generalising the first one, consider the family of mechanical systems with force function of the form 
$$
F(t,x,v)=k(t)\,v+\f(t,x),
$$
with $k(t)$ and $\f(t,x)$ being   arbitrary functions. In this case, for $\Gamma$ given by (\ref{mech}),  div$\,\G=k(t)$, and then forces of this type  admit as solution of equation (\ref{muforce}) the multiplier 
\be
\mu(t)=\exp\left(-{\int^t}k(z)\,dz\right),
\label{mut}
\ee
as pointed out above (\S 4.1). A solution of the system  (\ref{myH}) is 
\ben
m_x=-\mu(t)v,\quad m_v=0,\quad H=\mu(t)\left(\frac{v^2}{2}-\int^x\f(t,z)\,dz\right).
\een
Then, the affine Lagrangian is
\be
L(t,x,v,\dot x,\dot v)=\exp\left({-{\int^tk(z)\,dz}}\right)\left(-v\,\dot x+\frac{v^2}{2}-\int^x\f(t,z)\,dz\right).
\ee

Choosing $q=x$ and $p=v\exp\left(-{\displaystyle \int^t}k(z)\,dz\right)$ as canonical conjugate coordinates, the Hamiltonian function is
$$
H(t,q,p)=\frac{p^2}{2}\exp\left({{\int^t}k(z)\,dz}\right)-\exp\left({-{ \int^t}k(z)\,dz}\right)\int^q\f(t,z)\,dz.
$$
 
As a particular instance we can consider a damped harmonic oscillator $\ddot x=-\w^2\,x-2b\,\dot x$, with $b>0,\,\w>0$, for which  the affine Lagrangian
\be
\label{lho}
L(t,x,v,\dot x,\dot v)=e^{2bt}\left(-v\,\dot x+\frac{1}{2}(v^2+ \w^2x^2)\right)
\ee
is obtained. The Lagrangian energy   $E_L=-\dfrac 12e^{2bt}(v^2+\w^2x^2)$ is not conserved. Passing to the Hamiltonian description, $\w_H=d\l= dx\wedge d(ve^{2bt})+dH\wedge dt$, so that a set of canonical variables is provided by $q=x$ and $p=ve^{2bt}$, with Hamiltonian function $H(t,q,p)=\dfrac 12(p^2e^{-2bt}+\w^2e^{2bt}q^2)$, which corresponds to  Bateman \cite{bateman} and Caldirola \cite{Cal41} formulations of dissipative harmonic oscillator (see also \cite{VP92}).

Incidentally, this result can be generalised for the 3-dimensional isotropic harmonic  oscillator with a linear damping, described by the equation
\be
\ddot{\bf r}+2b\,\dot{\bf r}+\w^2{\bf r}={\bf 0},
\ee
and which is equivalent to three independent equally damped  harmonic oscillators  (the normal modes). The Lagrangian is now
\be
L=e^{2bt}\left(-{\bf v}\cdot\dot{\bf r}+\frac{{\bf v}^2}{2}+\frac{1}{2}\w^2{\bf r}^2\right)\,.
\ee
This very simple derivation is to be compared with others previous arguments as, for instance, the one given by Havas \cite{ha73}. 

The Lane-Emden equation is also an interesting example of this type of forces, appearing, among other sources, from the Poisson equation for the gravitational field of fluids with a spherical symmetry in a hydrostatic equilibrium and that appears in many other different contexts. For instance, it is used in astrophysics \cite{cha42} for modeling  star as a symmetric star of gases in equilibrium under its own weight. In the case of a polytropic state equilibrium, the Lane-Emden equation is
\be
\label{lemq}
\ddot x+\frac{2}{t}\dot x+x^n=0,
\ee
and therefore it admits the Lagrangian
\be
\label{llemp}
L(t,x,v,\dot x,\dot v)=t^2\left(-v\,\dot x+\frac{v^2}{2}+\frac{x^{n+1}}{n+1}\right).
\ee

This result can be extended to its obvious generalisation, the Emden equations $\ddot x+a(t)\,\dot x+b(t)\,x^n=0$, as  for instance, the Bessel equation.

The Hamiltonian formulation is achieved by choosing  $q=x$, $p=v\,t^2$ as canonical variables, and the Hamiltonian function is 
$$H(t,q,p)=\frac{p^2}{2t^2}+t^2\frac{q^{n+1}}{n+1},$$ the same result as the one given in \cite{TdC09}.

 \medskip
 
Further generalisation of the previous cases leads us to consider  forces $F$  of the form 
\begin{equation}
F(t,x,v)=A(t,x)+B(t,x)v+C(t,x)v^2,\label{Fquadrv}
\end{equation}
which admit
a $v$-independent multiplier $\mu(t,x)$ when 
 there exists a function $\f(t,x)$ such that $d\f=B\, dt+2C\,dx$,  and then  the multiplier and the force are just $\mu=e^{-\f}$ and  $F=A(t,x)+(\partial\f/\partial t)v+(1/2)(\partial \f/\partial x)v^2$, a result   already obtained by Jacobi, as indicated in \cite{nu10}.

With the general prescription,  we find that an affine Lagrangian for this family of forces is
\be
\label{lfmu(t,x)}
L(t,x,v,\dot x,\dot v)=e^{-\f(t,x)}\left(\frac{v^2}{2}-v\,\dot x\right)-\int^xe^{-\f(t,z)}A(t,z)\,dz.
\ee

As an application of this result, let us consider the  second-order differential equation  
\be
\ddot x=\frac{3\dot x^2}{x}+\frac{\dot x}{t},\quad x\ne0,
\ee
derived by Buchdahl in General Relativity \cite{bu64}, for which $A=0$, $\f(t,x)=\log(tx^6)$ and $ \mu=1/(tx^6)$. Consequently,
\be
L(t,x,v,\dot x,\dot v)=\frac1{t\,x^6}\left(\frac{v^2}{2}-v\,\dot x\right).
\ee

Additionally,  we can reverse the meaning of the equation (\ref{muforce}) by fixing  the form of the  the Jacobi multipliers and asking for  the family of forces whose associated second-order differential equations admit such  functions as a Jacobi multipliers.

The trivial case of the multiplier being a real constant, for instance $\mu=1$, demands a $v$-independent force $F=\f(t,x)$ and it has been studied as a first example. The case in which  the Jacobi multiplier $\mu$ is a function  only of $t$ was studied in 3) of Subsection 4.1, and then the equation (\ref{muforce}) reduces to 
 $$
 \frac {d\log \mu}{dt}+\pd Fv=0,
 $$
 which shows that  
 \be
F (t, x, v) = k(t) v + \varphi (t, x), \label{fmu(t)}
\ee
with  $\varphi$  being an arbitrary function and $k(t)=-d(\log \mu)/dt$. Conversely, a Jacobi multiplier for the force (\ref{fmu(t)})  is given by $\mu(t) =\exp\left(-{\displaystyle \int^t}k(\zeta)d\zeta\right)$. This is the second case we have studied above.

When we consider the  particular case for which the  function $\mu$   only depends on $x$ and look for the forces $F$ admitting such a multiplier,  as  the equation (\ref{muforce}) reduces to 
 $$v\frac {d\log \mu}{dx}+\pd Fv=0,
 $$
we see  from here   that $F$ must be of the form 
 \be
\label{fmu(x)}
F(t,x,v)=k(x)\,{v^2}+\varphi(t,x),
\ee
where $\varphi $ is an arbitrary function and $2k(x)=-d(\log\mu)/dx$.   The multiplier  for a force like (\ref{fmu(x)})  is of the form 
\be
\mu(x)=\exp\left(-{ 2\int^x k(\zeta)\, d\zeta}\right).
\label{mu(x)}
\ee
This is a particular case of those of (\ref{Fquadrv}) that we have studied before.

The case of a multiplier  that depends only on the velocity $v$ can be  analysed along the same lines. In this case 
the equation (\ref{muforce}) is
 $$
 F\frac {d\log \mu}{dv}+\pd Fv=0,
 $$
 or written in a different way,
 $$
 \pd{ }v(\log \mu+\log F)=0,
 $$
and then the force $F$ which admits such a multiplier $\mu(v)$ must be of the form
\be
F(t,x,v)=\f(t,x)\,\Phi(v),
\ee
with $\Phi(v)=1/\mu(v)$ and $\varphi(t,x)$ an arbitrary function.

For  a $v$-independent multiplier, $\mu=\mu(t,x)$, the equation (\ref{muforce}) is
$$
\pd{\log \mu}t+v\pd{\log \mu}x+\pd{F}v=0,
$$
and one obtains that the forces $F$ admitting such a multiplier are those of the form
\be
\label{fmu(t,x)}
F(t,x,v)=A(t,x)+B(t,x)v+C(t,x)v^2,
\ee
where $A(t,x)$ is an arbitrary function, and 
$$B(t,x)=-\pd{ \log\mu}t, \qquad C(t,x)=-\frac 12\pd{ \log\mu} x.$$ That is, $B$ and $C$ are  functions satisfying the relation $\partial_x B=2\partial_t C$, which is equivalent to  say that the 1-form $B\,dt+2C\,dx$ is closed (i.e. locally  exact), and then there exists a locally defined function $\f(t,x)$ such that $d\f=B\,dt+2C\,dx$. Therefore this is the case studied in (\ref{Fquadrv}), and then a Jacobi  multiplier is $\mu=e^{-\f}$ and the force, $$F=A(t,x)+\pd \f t \,v+\frac 12 \pd  \f x\, v^2,$$ a result   already obtained by Jacobi, as indicated in \cite{nu10}.

  The same procedure can be applied to other cases, as, for instance,  a  multiplier of the form $\mu(t,v)=a(t)b(v)$,  for which the equation (\ref{muforce}) is
 $$
 \frac{\dot a}a +F\frac{b'}b+\pd{F}v=0,
$$
a nonhomogeneous  linear differential equation which can be integrated and we find the  family of forces
\be
\label{fa(t)b(v)}
F(t,x,v)=B(v)\left(\f(t,x)+A(t)\int^v\frac{dz}{B(z)}\right),
\ee
with $A(t)=-\dot a/a$, $B(v)=1/b(v)$, and $\f$ being an arbitrary function of $t$ and $x$.

\subsection{Biological systems}
\indent

We can  give a mathematical formulation in order to the study the evolution of certain biological populations of two competing species   as a system of   
first-order differential equations in the plane, so that they admit Lagrangian and Hamiltonian descriptions. 
This can now be illustrated by means of  two examples: a  generalisation of the Lotka-Volterra model \cite{VV31} and a host-parasite model \cite{Tru74}.

 \medskip
  
(1)  Consider first    the 2-dimensional system of nonlinear  differential equations
 \be
  \left\{\begin{split}
  \,&\dot x=x(A+Bx+Cy)\\ 
  &\dot y=y(K+Mx+Ny)
  \end{split}\right. \quad ,
  \label{LV}
  \ee
 where $A,B,C, K,\dots$,  are constant parameters 
 and the domain of interest is the region of $\mathbb{R}^2$  where $x>0$ and $y>0$. The model described by (\ref{LV}) 
it is frequently found in analysing the behaviour of biological interacting species, the parameters $A,B,\dots$, being the growth and interaction (encounters) coefficients; for instance, 
in the Lotka-Volterra model for the evolution of a predator-prey system assuming logistic growth for preys ($x$) when there are no predators ($y$), the coefficients  $A$ and $M$ are positive, $B,\,C$ and $K$ negative,  and $N=0$ \cite{jo10}. The equations (\ref{LV}) are of the first-order already  \cite{jo10} and the associated  vector field is
\be
 \G=\pd{}t+x(A+Bx+Cy)\pd{}{x}+y(K+Mx+Ny)\pd{}y\,.
\ee
Therefore,  as $\dg \Gamma=A+2B\, x+C\,y+K+2N\, y+M\, x$, the equation  corresponding to (\ref{mun}) for the Jacobi multiplier is
\ben
\pd{\mu}{t}+x(A+Bx+Cy)\pd{\mu}{x}+y(K+Mx+Ny)\pd{\mu}{y}+\mu\big(A+K+(2B+M)x+(C+2N)y\big)=0.
\een
A simple solution $\mu$, valid no matter the value of the constants $A,B,\dots$, can be easily  achieved by assuming that the multiplier is of the form
\be
\mu(t,x,y)=e^{rt}x^p y^q,
\label{mLV}
\ee
that is,  such that $\partial_t\mu=r\,\mu$, $x\partial _x\mu=p\,\mu$ and $y\partial_y\mu=q\,\mu$. The needed exponents $p,\,q$ and $r$ are solution of the linear system
\be
  \left\{\begin{split}
  \,&Bp+Mq+2B+M=0\\ 
  &Cp+Nq+2N+C=0\\
  &r+A(p+1)+K(q+1)=0\,
  \end{split}\right. \quad .
  \label{expon}
  \ee
When $\Delta=BN-CM\ne0$ the solution is unique and with the resulting multiplier we get, by solving the system (\ref{myH}),  the particular solution 
\be
m_x=-\frac{e^{rt}x^p y^{q+1}}{q+1},\quad 
m_y=0,\quad 
H=\frac{e^{rt}x^{p+1}y^{q+1}}{q+1}\left(A+Bx+C\frac{q+1}{q+2}\,y\right)\,, \label{1fglv}
\ee
a solution that obviously is only valid when $(q+1)(q+2)\ne0$.

In this way we obtain an affine Lagrangian for the 2-dimensional  system (\ref{LV}):
\be
\label{LLV}
L=\frac{e^{rt}x^{p+1}y^{q+1}}{q+1}\left(-\frac{v_x}x+A+Bx+C\frac{q+1}{q+2}\,y\right)\,.
\ee
The connection between the parameters $A,B,C, K, M$ and $N$ and the exponents $r,p,q$ is given by (\ref{expon}).

Note that the classical Lotka--Volterra system, that is, the predator-prey system with exponential growth \cite{jo10}, corresponds to put $B=N=0$ in (\ref{LV}), then $p+1=q+1=0$ and $r=0$. The equation (\ref{mLV}) gives then  the following multiplier 
\be
\mu=\frac 1{xy}\,,
\ee
but now  (\ref{1fglv}) and  (\ref{LLV}) are not   valid  and it is necessary to analyse this case separately. The right expressions for $m_x$ and $H$ are now
\ben
m_x=-\frac{\log y}{x}\,,\quad H=-K\log x-Mx+A\log y+Cy\,.
\een
In this way,  we obtain a Lagrangian for the general 2-dimensional Lotka--Volterra  system, by taking $m_y=0$ once again:
\be
L(t,x,y,v_x,v_y)=-\frac{\log y}{x}v_x-K\log x-Mx+A\log y+Cy\,.
\ee
The Lagrangian energy, $E_L=-H=K\log x+Mx-A\log y-Cy$, is conserved.
There are other solutions with the same multiplier but every one differing from each other by a total time derivative   \cite{fn98}. 
 
As for the other previous examples, we can  give a Hamiltonian formulation the Lotka-Volterra system: the canonical conjugate variables are $\tilde q= x$ and $\tilde p=(\log y)/x$, 
with Hamiltonian function $H(\tilde q,\tilde p)=-K\log \tilde q-M\tilde q+A\tilde p\,\tilde q+Ce^{\tilde p\,\tilde q}$.

\medskip

(2) Another interesting example is the one given by the nonlinear system of differential equations
\[
\label{logLV}
\left\{\begin{split}
&\dot x=x(A-B\,y)\\  
&\displaystyle\dot y=y\left(C-D\,\frac yx\right)
\end{split}\right. \quad,
\]
 known as the ``host-parasite'' model \cite{Tru74}. The region of interest is again $x>0,\, y>0$, and $A,B,C$ and $D$ are positive real constants. The relevant vector field is 
 $$
 \Gamma=\pd{}t+x(A-B\,y)\pd{}x +y\left(C-D\,\frac yx\right)\pd{}y,
 $$
 and as  $\dg \Gamma=A-B\,y+C-2D\dfrac{y}x$, the equation (\ref{mun}) for multiplier results
\ban
\pd{\mu}t+x(A-B\,y)\pd{\mu}x
+y\left(C-D\,\frac yx\right)\pd{\mu}{y}+\mu\left(A-B\,y+C-2D\, \frac{y}x\right)=0.
\ean
As in the preceding example, we can look for a solution like $\mu=e^{rt}x^py^q$, and  the resulting multiplier is, up to a constant factor,
\[
\mu=\frac{e^{Ct}}{xy^2}\,.
\]
Writing  the equations (\ref{myH}) we find for the coefficients $m$ and the function $H$ 
\[
m_x=\frac{e^{Ct}}{xy},\quad m_y=0,\quad H=-e^{Ct}\left(\frac Dx+\frac Ay+B\log y\right),
\]
so that an affine Lagrangian is
\[
L(t,x,y,v_x,v_y)=e^{Ct}\left(\frac{v_x}{xy}-\frac Dx-\frac Ay-B\log y\right).
\] 
Other Lagrangians with the same multiplier can be easily found, all of them being gauge-equivalent. For example,  the affine Lagrangian we have obtained differs from the one given in \cite{Tru74}, but they are gauge-equivalent.

With canonical variables $q=x$ and $p=-m_x=-e^{Ct}/(xy)$  and   Hamiltonian 
\[
H=e^{Ct}\left(-\frac Dq+Ae^{-Ct}pq-BCt+B\log|pq|\right),
\]
the dynamical equations are recovered as $\dot q=\partial_p H,\,\dot p=-\partial_q H$.
\section{Summary and outlook}
\indent 

 We have analysed the 2-dimensional inverse problem for first-order systems and devised a method to construct an affine Lagrangian for every system of two first-order ordinary 
 differential equations. A time-dependent Hamiltonian formulation for
the dynamical system $\G\in\X(\R\x\R^2)$ 
describing the dynamics of our Lagrangian (\ref{ala}) is also studied  in the regular case $\mu(t,x,y)\ne0$.  The construction of  the affine  Lagrangian is based on the knowledge of a Jacobi multiplier for the given system of differential equations, 
 with respect to the usual volume form. Of course the Lagrangian so obtained is singular.  The method is 
 particularly suitable for the case of the first-order systems equivalent to a single second-order differential equation, typically those describing 
 the evolution of mechanical systems, and the theory has been illustrated with several examples, depending on the explicit form of the forces. We have  also  reversed the method and found, in the case of mechanical systems, the family of forces admitting a given function as a multiplier. This point of view is fruitful and produces new examples of affine Lagrangians.
 
  It is also remarkable that the method  is equally applicable to systems of  first-order differential equations used in the mathematical modeling of interacting biological species \cite{Tru74} and several examples, as  a  generalisation of the Lotka-Volterra model and a host-parasite model,  have explicitly been developed in Subsection 5.2.
 
We think that the generalisation  to higher dimensional cases would be of interest.  We give only a summary of the theoretical results generalising the 2-dimensional case we have treated, and mention a pair of 4-dimensional examples.

An affine Lagrangian  on a $2n$-dimensional manifold $M$ (for instance, $\R^{2n}$) is a differentiable function  $L$ locally given by
\be
\label{nL}
L=\sum_{i=1}^n m_i(t,x)\, v^i+H(t,x),\quad x=(x^1,\ldots,x^n),
\ee
where $(t,x^i)$, $i=1,\dots,2n$, are a system of local coordinates  on $\R\x M$ and $(t,x^i,v^i)$ the corresponding fibred ones  on $\R\x TM$,  and $m_i, H\in C^\infty(\R\x M)$. 

As in the 2-dimensional case, the function 
$L$ is obtained by contraction of the 1-form $\l={\displaystyle \sum_{i=1}^n} m_i\,dx^i+H\, dt$ with the time-derivative operator 
$\T=\partial_t+{\displaystyle \sum_{i=1}^n} v^i\partial_{x^i}$. The corresponding system of Euler-Lagrange equations is the system of first-order differential equations
\be
\label{beq}
\sum_{j=1}^n \mu_{ij}v^j=\wr_i\,,\quad i=1,\ldots,n,
\ee
where $\mu_{ij}=\partial_{x^i} {m_j} -\partial_{x^j} {m_i}$ and $\wr_i=\partial_t {m_i}-\partial_{x^i} {H}$; 
they are the obvious generalisations of (\ref{ec1}) and (\ref{mu}) and we are interested only in the `regular' case, namely, when the matrix  $\mu$ is regular. The equations (\ref{beq}) are sometimes known as the 
Birkhoff's equations, see e.g. \cite{MGX06, sa83}.

The inverse problem is formulated in the same terms as before:  given the vector field $\G=\partial_t+
X^i\partial_{x^i}\in\X(\R\x M)$, is there an affine Lagrangian (\ref{nL}) whose system of Euler--Lagrange equations is equivalent
 to the first-order system $\dot x^i=X^i(t,x)$ for $i=1,\ldots,n$? The necessary and sufficient condition is the existence of a non-degenerate exact invariant integral 2-form $\a$ for $\G$, that is, a non-degenerate 2-form $\a\in\bigwedge\!^2(\R\x M)$ such that $d\a=0$ and $i(\G)\a=0$.

If in the local coordinates $(t,x)$ the 2-form $\a$ is 
$$
\a=\frac 12\sum_{i,j=1}^nA_{ij}\,dx^i\wedge dx^j+\sum_{i=1}^n B_i\,dt\wedge dx^i,
$$ 
with the matrix $(A_{ij})$ being skew-symmetric, 
these conditions are expressed by means of $\det A\ne 0$, $B_i={\displaystyle\sum_{j=1}^n}A_{ij}X^j$, for $i=1,\ldots,n$, and the following system of linear PDE that the $A$-coefficients must satisfy:
\be
\label{theA}
\pd{A_{ij}}{t}+\sum_{k=1}^n\left(X^k\pd{A_{ij}}{x^k}+A_{kj}\pd{X^k}{x^i}+A_{ik}\pd{X^k}{x^j}\right)=0\,, \quad i,j=1,\dots,2n.
\ee

The main problem is now to find a particular solution of (\ref{theA}) for the $n(2n-1)$ skew-symmetric coefficients $A_{ij}$, a problem that might  be a tough task in practical cases. After that, it is necessary to  find the 1-form $\l$ that integrates $\a$ and construct the affine
 Lagrangian $L=i_\T\l$.  For instance, the system of two second-order differential equations $\ddot x=-\dot y$, $\ddot y=-y$, a non-Lagrangian example in Douglas' classification \cite{do41}, is equivalently represented in the variables $x_1=x$, $x_2=y$, $x_3=v_x$, $x_4=v_y$ by the linear first-order system  on $\R^4$ 
\be
\label{2example}
\left\{
\begin{split}
\,&\dot x_1=x_3\\
&\dot x_2=x_4\\
&\dot x_3=-x_4\\
&\dot x_4=-x_2
\end{split}
\right.\quad .
\ee

Solving (\ref{theA}) for (\ref{2example}) requires searching for six $A$-coefficients. A simple solution  all of them   being constant does exist and yields various affine Lagrangians, as, for instance,
\be
L=(x_2+x_3)\dot x_1-x_3\dot x_4+\frac{1}{2}(-x_3^2+x_4^2-2x_2x_3),
\ee
which is gauge-equivalent to the Hojman-Urrutia Lagrangian \cite{ho91,ho81}, obtained by previous  integration of the equation. There are other  non gauge-equivalent possibilities, as 
\be
L_1=(x_2+x_3)\dot  x_1+(x_2-x_3)\dot x_4+\frac{1}{2}(x_2^2-x_3^2+2x_4^2-2x_2x_3).
\ee
Clearly, there is no function $f$ such that  $L_1-L= df/dt$.

Another example we can  consider  is $\ddot x=y^2, \ddot y=x^2$. We can check that  the equivalent first-order system  $\dot x_1=x_3$,  $\dot x_2=x_4  $,   $\dot x_3=x_2^2  $,   $\dot x_4=x^2_1$ admits an affine Lagrangian  description with   $L=x_4\dot x_1+x_3\dot x_2+(x_1^3+x_2^3)/3-x_3x_4$.

As a final example let us consider briefly a more involved example: the 4-dimensional Lotka-Volterra system given by the system of differential equations
\begin{equation}
\left\{\begin{split}
\,&\dot x_1=x_1(-1+x_2)\\
&\dot x_2=x_2(1-x_1+ax_3)\\
&\dot x_3=x_3(-1-ax_2+x_4)\\
&\dot x_4=x_4(1-x_3)
\end{split}\right.,\label{duarte}
\end{equation}
with $a$ being a constant parameter,  proposed in \cite{DFO}. A particular solution of equations (\ref{theA}) for the six  $A$-coefficients appearing in the expression of the 2-form $\alpha$   is $A_{12}=(x_1x_2)^{-1},\,A_{14}=a(x_1x_4)^{-1},\, A_{34}=(x_3x_4)^{-1},\, A_{13}=A_{23}=A_{24}=0$,
 while $$B_1={\displaystyle{\sum_{j=1}^3}}A_{1j}X^j=-1+\frac{1+a}{x_1}, \qquad B_2={\displaystyle{\sum_{j=1}^3}}A_{2j}X^j=-1+\frac 1{x_2},$$,   $$B_3={\displaystyle{\sum_{j=1}^3}}A_{3j}X^j=-1+\frac 1{x_3},  \quad B_4={\displaystyle{\sum_{j=1}^3}}A_{4j}X^j=-1+
 \frac{1+a}{x_4},$$
 to which it corresponds the affine Lagrangian
$$
L=\frac{\log x_1}{x_2}v_2-\frac{\log x_4}{x_3}v_3+\frac{\log x_1}{x_4}v_4+x_1+x_2+x_3+x_4-\log(x_1^{1+a}\,x_2\,x_3\,x_4^{1+a}).
$$
It is easy to check that the corresponding Euler-Lagrange equations are equivalent to (\ref{duarte}).

\vspace{20pt}

{\bf Author contributions:} {Investigation, J.F.C. and J.F.N. Both authors have read and agreed
to the published version of the manuscript}

\medskip

{\bf Funding:} {This research was supported by  the Spanish Ministerio de
Ciencia, Innovaci\'on y Universidades project
PGC2018-098265-B-C31}
 
\medskip

{\bf Conflicts of interest:} {The authors declare no conflict of interest}

\end{document}